\documentclass[sort&compress,twocolumn,10pt,5p]{elsarticle}
\usepackage{atlasphysics}

\usepackage{subfigure}
\usepackage{mathrsfs}
\usepackage{graphicx}
\usepackage{multirow}
\usepackage{rotating}
\usepackage{xspace}
\usepackage{slashed}
\usepackage{url}
\usepackage[hyperindex,breaklinks]{hyperref} 

\journal{Physics Letters B}

\def\coupling{\ensuremath{k/\overline{M}_{Pl}}}
\def\mpl{\ensuremath{\overline{M}_{Pl}}}

\begin{document}
\begin{frontmatter}

\title{\vspace{-0.5cm} \flushright{\normalsize{CERN-PH-EP-2011-189}} \\ \center{Search for Extra Dimensions Using Diphoton Events in \\ 7~TeV Proton-Proton Collisions with the ATLAS Detector}}

\author{The ATLAS Collaboration}

\begin{abstract}

Using data 
recorded in 2011 with the ATLAS detector at the Large Hadron Collider, a 
search for evidence of extra spatial dimensions has been performed
through an analysis  of the diphoton final state. 
The analysis uses data corresponding to an integrated luminosity 
of $2.12$~fb$^{-1}$ of  $\sqrt s= 7$~TeV proton-proton collisions.
The diphoton invariant mass ($m_{\gamma\gamma}$) spectrum is observed to be in good agreement with
the expected Standard Model background. 
In the large extra dimension scenario of Arkani-Hamed, Dimopoulos and Dvali, the results
provide 95\% CL lower limits on the fundamental Planck scale between 2.27 and 3.53~TeV, 
depending on the number of extra dimensions and the theoretical formalism used. 
The results also set 95\% CL lower limits on the lightest Randall-Sundrum graviton mass 
of between 0.79 and 1.85~TeV, for values of the dimensionless coupling 
 \coupling \  varying from 0.01 to 0.1.  Combining with previously
 published ATLAS results from the dielectron and dimuon final states,  the 95\% CL lower limit on the 
 Randall-Sundrum graviton mass for $\coupling  = 0.01 \ (0.1)$ is $0.80 \ (1.95)$~TeV. 
\end{abstract}
\end{frontmatter}

\section{Introduction}
The enormous difference between the Planck scale and the 
electroweak scale is known as the hierarchy problem.  
A prominent class of new physics models addresses the hierarchy problem
through the existence of extra spatial dimensions.
In this paper, we search for evidence of extra dimensions within the context of the models of
Arkani-Hamed, Dimopoulos, and Dvali (ADD)~\cite{ADD} and of Randall and Sundrum (RS)~\cite{RSmodel}.
In these models, gravity can propagate in the higher-dimensional 
bulk, giving rise to a so-called Kaluza-Klein (KK) tower of massive spin-2 
graviton excitations (KK gravitons, $G$).  Due to their couplings to Standard Model
(SM) particle-antiparticle pairs, KK gravitons can be investigated in proton-proton ($pp$)
collisions at the Large Hadron Collider (LHC) via a variety of processes, including 
virtual graviton exchange as well as direct
graviton production through gluon-gluon fusion or quark-antiquark annihilation.

The ADD model~\cite{ADD}
postulates the existence of $n$  flat additional spatial dimensions compactified with radius $R$,
in which only gravity propagates. The fundamental Planck scale in the $(4+n)$-dimensional 
spacetime, $M_{D}$, is related to the apparent scale $M_{Pl}$ by Gauss' law: 
$\mpl^{2} = M_{D}^{n+2}R^{n}$, where $\mpl = M_{Pl}/\sqrt{8\pi}$ is the reduced Planck scale.
The mass splitting between subsequent KK states is of order $1/R$.
In the ADD model, resolving the hierarchy problem requires typically small values of $1/R$, giving rise to an
almost continuous spectrum of KK graviton states.

While processes involving direct graviton emission depend on $M_D$, effects involving
virtual gravitons depend on the ultraviolet cutoff of the KK spectrum, denoted $M_{S}$.  
The effects of the extra dimensions  are typically parametrized by $\eta_{G} = \mathcal{F}/M_S^4$, where
$\eta_G$ describes the strength of gravity in the presence of the extra dimensions and 
$\mathcal{F}$ is a dimensionless parameter of order unity reflecting the dependence of virtual 
KK graviton exchange on  the number of extra dimensions. Several theoretical 
formalisms exist in the literature, using different definitions 
of $\mathcal{F}$ and, consequently, of $M_{S}$:

\begin{equation}
\label{eq:ADD_GRW}
\mathcal{F}  =  1 ,    \textrm{ (GRW)~\cite{GRW}}; \\
\end{equation}

\begin{equation}
\mathcal{F} = \left\{ \begin{array}{ll}
\textrm{log}\left(\frac{M_{S}^{2}}{\hat s}\right) & \textrm{$n = 2$} \\
\frac{2}{n-2} & \textrm{$n > 2$}
\end{array} \right.
, \textrm{ (HLZ)~\cite{HLZ};} \\
\label{eq:ADD_HLZ}
\end{equation}

\begin{equation}
\label{eq:ADD_Hew}
\mathcal{F}  =   \pm\frac{2}{\pi} ,    \textrm{ (Hewett)~\cite{Hewett}};
\end{equation}

\noindent where $\sqrt{\hat s}$
is the center-of-mass energy of the parton-parton collision.
Effects due to ADD graviton exchange would be
evidenced by a non-resonant deviation from the SM background expectation. 
Collider searches for ADD virtual 
graviton effects have been performed at HERA~\cite{HERAADD}, LEP~\cite{LEPADD},
the Tevatron~\cite{TevatronADD}, and the LHC~\cite{CMSADD,CMSggnew}. Recent diphoton results from
CMS are the most restrictive so far,
setting limits on $M_S$ in the range of 2.3 to 3.8~TeV~\cite{CMSggnew}. 

The RS model~\cite{RSmodel} 
posits the existence of a fifth dimension with ``warped'' geometry, bounded
by two (3+1)-dimensional branes, with the SM fields localized on the so-called TeV brane and gravity originating
on the other, dubbed the Planck brane, but capable of propagating in the bulk.  
Mass scales on the TeV brane, such as the Planck mass describing the observed
strength of gravity, correspond to mass scales on the Planck brane as given by
$M_D = M_{Pl} e^{-k\pi r_c}$, where $k$ and $r_c$ are the curvature
scale and compactification radius of the extra dimension, respectively.
The observed hierarchy of scales can therefore be naturally reproduced
in this model, if $kr_c \approx 12$~\cite{coupling}. KK gravitons in this model would have a
mass splitting of order 1~TeV and would appear as new resonances.  
The phenomenology can be described in 
terms of the mass of the lightest KK graviton excitation ($m_G$) and the dimensionless coupling
to the SM fields, \coupling. It is theoretically preferred~\cite{coupling}
for \coupling \ to have a value in the range from 0.01 to 0.1. 
The most stringent experimental limits on RS gravitons are from the LHC. 
For \coupling = 0.1, $\sim$1~fb$^{-1}$
ATLAS results from $G \rightarrow ee/\mu\mu$
exclude gravitons below 1.63~TeV~\cite{zprime}, assuming
leading order (LO) cross section predictions, and a recent 2.2~fb$^{-1}$ $G \rightarrow \gamma\gamma$ result
from CMS excludes gravitons below 1.84~TeV~\cite{CMSggnew}, using
next-to-leading order (NLO) cross section values.  These results have
surpassed the limits from searches at the Tevatron~\cite{TevatronRS} and
earlier searches at the LHC~\cite{CMSRSeemm}.

The diphoton final state provides a sensitive channel for this search due to the clean 
experimental signature, excellent diphoton mass resolution, and modest backgrounds, 
as well as a branching ratio for graviton decay to diphotons that is twice the value of that
for graviton decay to any individual charged-lepton pair.
In this letter, we report on a search in the diphoton final state for evidence of extra dimensions,	
using a data sample corresponding to an integrated luminosity 
of $2.12$~fb$^{-1}$ of  $\sqrt s= 7$~TeV $pp$ collisions,
recorded during 2011 with the ATLAS detector at the LHC.
The measurement of the diphoton invariant mass spectrum is interpreted in both the ADD and RS scenarios.
  
\section{The ATLAS Detector}

The ATLAS detector~\cite{DetectorPaper} is a multipurpose particle
physics instrument with a forward-backward symmetric cylindrical
geometry and near 4$\pi$ solid angle coverage\footnote{ATLAS 
uses a right-handed coordinate system with its origin at the nominal
interaction point in the centre of the detector and the $z$-axis along the
beam pipe. Cylindrical coordinates $(r,\phi)$ are used in the transverse
plane, $\phi$ being the azimuthal angle around the beam pipe. The pseudorapidity $\eta$ is
defined in terms of the polar angle $\theta$ by $\eta=-\ln\tan(\theta/2)$.}.
Closest to the beamline are tracking detectors to measure the trajectories of
charged particles, including layers of
silicon-based detectors as well as a transition radiation tracker using straw-tube 
technology. The tracker is surrounded by a thin solenoid that provides
a 2~T magnetic field for momentum measurements. The solenoid is surrounded by
a hermetic calorimeter system, which is particularly important for this analysis.  
A system of liquid-argon (LAr) sampling calorimeters is divided into
a central barrel calorimeter and two endcap calorimeters, each housed in a
separate cryostat. Fine-grained LAr electromagnetic (EM) calorimeters, segmented
in three longitudinal layers, are used to precisely
measure the energies of electrons, positrons and photons
for $\left|\eta\right| < 3.2$. Most of the EM shower energy is 
collected in the second layer, which has a granularity of $\Delta \eta \times
\Delta \phi = 0.025 \times 0.025$. The first layer is segmented into eight strips
per middle-layer cell in the $\eta $ direction,  extending over four middle-layer cells in $\phi $, designed
to separate photons from $\pi^0$ mesons. 
A presampler, covering $\left|\eta\right| < 1.81$,  is used to correct for
energy lost upstream of the calorimeter. The regions spanning
$1.5 < \left|\eta\right| < 4.9$ are instrumented with LAr calorimetry also
for hadronic measurements, while an iron-scintillator
tile calorimeter provides hadronic coverage in the range $\left|\eta\right| < 1.7$.
A muon spectrometer consisting of three superconducting toroidal magnet systems, tracking chambers, 
and detectors for triggering lies outside the calorimeter system.

\section{Trigger and Data Selection}

The analysis uses data collected between March and September 2011 
during stable beam periods of 7~TeV $pp$ collisions.
Selected events had to satisfy a trigger requiring at least two photon candidates
with transverse energy $\ET^\gamma > 20$~GeV and satisfying  a 
set of requirements, referred to as the ``loose'' photon definition~\cite{ATLASphotons},
which includes requirements on the leakage of energy into the hadronic calorimeter as well as on
variables that require the transverse width of the shower, measured in the second 
EM calorimeter layer, be consistent with the narrow width expected for an EM shower. 
The loose  definition is designed to have high photon efficiency, albeit with reduced
background rejection. The trigger was essentially fully efficient for high mass diphoton
events passing the final selection requirements.

Events were required to have at least one primary collision vertex,
with at least three reconstructed tracks.  Selected events had to have at least 
two photon candidates, each with $\ET^{\gamma} > 25$~GeV and pseudorapidity
 $|\eta^{\gamma} | < 2.37$, with the exclusion of  $1.37 < |\eta^{\gamma} | < 1.52$,
the transition region between the barrel and endcap calorimeters. As described in
more detail in Ref.~\cite{ATLASphotons}, photon candidates included those
classified as unconverted photons, with no associated track, or
photons which converted to electron-positron pairs, with one or two associated tracks. 
The two photons were required to satisfy several quality criteria and to lie
outside detector regions where their energy was not measured in
an optimal way.
The two photon candidates each had to satisfy a set of stricter
requirements, referred to as the ``tight'' photon definition~\cite{ATLASphotons}, which 
included a more stringent selection on the shower width in the second
EM layer and additional requirements on the energy distribution in the
first EM calorimeter layer. The tight photon definition was designed to 
increase the purity of the photon selection sample by rejecting 
most of the remaining jet background, including jets with a
leading neutral hadron (mostly $\pi^0$ mesons) that decay to a
pair of collimated photons.

The isolation transverse energy $\ET^{iso}$ for each photon
was calculated~\cite{ATLASphotons} by summing over the cells of both the EM and hadronic
calorimeters that surround the photon candidate within an angular cone of
radius $\Delta R = \sqrt{(\eta - \eta^\gamma)^2 + (\phi - \phi^\gamma)^2} < 0.4$,
after removing a central core that contains most of the energy of the photon. 
To reduce the jet background further, an isolation
requirement was applied, requiring that each of the two leading photons
satisfied $\ET^{iso} < 5$~GeV. 
An out-of-core energy correction was applied, to make $\ET^{iso}$
essentially independent of $\ET^\gamma$. An ambient energy
correction, based on the measurement of low transverse momentum jets~\cite{ambientenergy},
was also applied, on an event-by-event basis, to remove the contributions
from the underlying event and from ``pileup'', which results from
the presence of multiple $pp$ collisions within the same or nearby bunch crossings.  

For events with more than two photon candiates passing all the selection
requirements, the two photons with the highest $\ET^\gamma$ values were considered. 
The diphoton invariant mass had to exceed 140~GeV. A total of
6846 events were selected.

\section{Monte Carlo Simulation Studies}

Monte Carlo (MC) simulations were performed to study the detector response
for various possible signal models, as well as to perform some SM background studies.
All MC events  were simulated~\cite{ATLASsim}
with the ATLAS detector simulation based on {\sc geant4}~\cite{geant4} and using ATLAS
parameter tunes~\cite{ATLAStune}, and were processed through the same reconstruction software
chain as used for the data. 
The MC events were reweighted to mimic the pileup conditions observed in the data.

SM diphoton production was simulated with  PYTHIA~\cite{pythia} version 6.424 and
 MRST2007LOMOD~\cite{MRST2007lomod} parton distribution functions (PDFs). The
PYTHIA events were reweighted as a function of diphoton invariant mass to the differential cross
section predicted by the NLO calculation of DIPHOX~\cite{diphox} version 1.3.2. The
reweighting factor varied from $\approx 1.6$ for a diphoton mass of 140~GeV, decreasing
smoothly to unity for large masses.  For the DIPHOX  calculation, the
renormalization scale and the initial and final factorization scales of the model
were all set to the diphoton mass.  The various scales were varied by a factor of two both up
and down, compared to this central value, to evaluate systematic uncertainties.
The PDFs were chosen 
following the recommendations of the PDF4LHC working group~\cite{pdf4lhc}, with 
MSTW2008 NLO PDFs~\cite{mstw2008}  used for the NLO predictions, and CTEQ6.6~\cite{cteq66}
and MRST2007LOMOD~\cite{MRST2007lomod} used for systematic comparisons.

SHERPA~\cite{sherpa} version 1.2.3 was used with CTEQ6L~\cite{cteq66} PDFs
to simulate the various ADD scenarios for a variety of $M_S$ values. 
Due to the interference between the SM and gravity-mediated contributions, 
it is necessary to simulate events according to the full differential cross section as a function
of the diphoton mass. A generator-level cut was applied to restrict the signal simulation to diphoton 
masses above 200~GeV.   The ADD MC samples were used to determine the
signal acceptance ($A$) and selection efficiency ($\epsilon$).  The acceptance, defined as
the percentage of diphoton signal events with the two highest $\ET$ photons passing
the applied $\ET^\gamma$ and $\eta^\gamma$ cuts, varied somewhat for the
various ADD implementations and fell from typical values of $\approx 20$\% for $M_S = 1.5$~TeV
down to $\approx 15$\% for $M_S = 3$~TeV, due mostly to the variations in the $\eta^\gamma$ distributions.  
The selection efficiency, for events within the detector acceptance, was found to be $\approx 70$\%.

RS model MC signal samples were produced using the  
implementation of the RS model in PYTHIA~\cite{pythia} version 6.424, which is fully specified by providing the values of
$m_G$ and  \coupling. MC signal samples were produced for a range of $m_G$ and \coupling \ values, using the 
MRST2007LOMOD~\cite{MRST2007lomod} PDFs. The products of
$A \times \epsilon$ for the RS signal models were in the range
$\approx (53-60)$\%, slowly rising with increasing graviton mass.
The reconstructed shape of the graviton resonance
was modeled by convolving the graviton Breit-Wigner lineshape
with a double-sided Crystal Ball (CB) function to describe the detector response.
The natural width of the Breit-Wigner was fixed according to the expected theoretical value, which
varies as the square of \coupling.  The values of the width increase, for $\coupling = 0.1$, from
$\approx 8$~GeV up to $\approx 30$~GeV for $m_G$ values from 800~GeV to 2200~GeV, respectively. 
The parameters of the CB function, which includes a Gaussian core to model the detector 
resolution matched to exponential functions on both sides to model the modest non-Gaussian tails,
were determined by fitting to the reconstructed MC signals.  The fitted values of $\sigma$ of the 
Gaussian core approached a value of $\approx 1$\% for high $m_G$ values, as expected 
given the current value of the constant term in the EM calorimeter energy resolution, and
were found to be independent of \coupling. The EM energy resolution has been verified in data using
$Z \rightarrow ee$ decays~\cite{egamma}, and MC used to describe the modest differences between
the response to photons versus electrons. The fitted values of the CB parameters varied smoothly with
$m_G$.  Fitting this mass dependence provided a signal parametrization that was used
to describe signals with any values of $m_G$ and \coupling.

\section{Background Evaluation}

The largest background for this analysis is the irreducible background
due to SM $\gamma\gamma$ production. The shape of the diphoton
invariant mass spectrum from this background
was estimated using MC, reweighting the PYTHIA  samples to the differential
cross section predictions of DIPHOX.  

Another significant background component is the reducible background 
that includes events in which one or both of the reconstructed photon candidates
result from a different physics object being misidentified as a photon. This
background is dominated by $\gamma$ + jet ($j$) and 
$jj$ events, with one or two jets faking photons, respectively.  Backgrounds
with electrons faking photons, such as the 
Drell-Yan production of electron-positron pairs as well as $W/Z + \gamma$
and $t \bar t$ processes, were verified using MC to be small 
after the event selection and were neglected.
Several background-enriched control samples were defined in order to determine
the shape of the reducible background using data-driven techniques.
In all control samples, the two photon candidates were required to pass the same isolation 
cut as for the signal selection, since removing the isolation requirement was seen to 
modify the diphoton mass spectrum. The first control sample contained
those events where one of the photon candidates passed the tight  requirement applied
for the signal selection.  However, the other photon candidate was required to fail the tight
photon identification definition, but to pass the loose  requirement; the latter restriction was
applied to avoid any trigger bias, as the trigger required two loose photons.   This sample is
enriched in $\gamma + j$ events, where the photon passed the tight requirement and a jet passed 
the loose one,
and also in $jj$ events where both photon candidates were due to jets.
A second control sample, dominated by $jj$ events, was similarly defined, but both photon
candidates were required to fail the tight photon identification while passing the loose definition.

The diphoton invariant mass distributions were compared for these control samples.  
To check for any kinematic bias, the control sample with one tight  and one loose
photon candidate was further divided, with the $\gamma j$
($j\gamma$) subsample being defined as the case with the  tight  photon being the 
photon candidate  with the highest (second highest) transverse energy. The
diphoton invariant mass distributions of all three control
subsamples were found to be consistent with each other, within statistical uncertainties.  
The sum of the control samples was used 
to provide the best estimate of the reducible background shape.
Variations among the subsamples were taken into account as a source of 
systematic uncertainty in the reducible background prediction.

The data control samples have relatively few events in the high diphoton 
mass signal region. It was therefore necessary to
extrapolate the reducible background shape to higher masses, which was done
by fitting with a smooth function of the form $f(x) = p_1 \times x^{p_2 + p_3 \log x}$,
where $x = m_{\gamma\gamma}$ and $p_i$ are the fit parameters.  
This functional form has been used in previous ATLAS
resonance searches~\cite{zprime,ATLASdijet}, and 
describes well the shape of the control data samples.  

The total background, calculated as the sum of the irreducible and reducible components,
was normalized to the number of data events in a low mass control region with diphoton
masses between 140 and 400~GeV, in which possible ADD and RS signals have been excluded
by previous searches.  The fraction of the total background in this region that
is due to the irreducible background is defined as the purity of the sample.  The purity ($p$) was
determined by three complementary methods.
The most precise measurement resulted from a method previously used in 
References~\cite{Hgg,SMdiphoton} that examines the 
 $\ET^{iso}$ values of the two  photon candidates.
Templates for the $\ET^{iso}$ distributions of true photons and of fake photons from
jets were both determined from the data.  
The shape for fake photons was found using a sample of photon candidates that failed at least one of a subset of several
of the selection requirements used for the tight  photon definition. 
The shape for photons was found
from the tight  photon sample, after subtracting the fake photon shape normalized to match
the number of candidates with large values (greater than $10$~GeV) of $\ET^{iso}$.  In addition, for $jj$ events, 
due to the observed significant ($\approx 20$\%) correlation between the $\ET^{iso}$ values of the 
two photon candidates, a two-dimensional template was formed using events in which 
both photon candidates failed the tight  identification. An extended maximum 
likelihood fit to the two-dimensional distribution formed from the $\ET^{iso}$ values of the two  
photon candidates was performed 
in order to extract the contributions from $\gamma\gamma$, $\gamma j$, $j \gamma$, and
$jj$ events. The fit was performed using the photon and fake photon $\ET^{iso}$ templates, 
as well as the two-dimensional $jj$ template.  The resultant value of the purity in
the low mass control region was $p = 71^{+5}_{-9}$\%.
The uncertainty was determined by varying the subset of tight selection criteria failed by fake
photon candidates, and then repeating the purity determination.
Cross checks using either the DIPHOX prediction for the
absolute normalization of the irreducible component, or fitting the shapes of the
irreducible and reducible backgrounds to the data in the low mass control region,
yielded consistent, but less precise results.  The result from the isolation method
was therefore used as the best estimate of the purity, and the total 
SM background prediction was set equal to the sum of the irreducible
and reducible components, weighted appropriately by this purity value
and normalized to data in the low mass control region.

\section{Systematic Uncertainties}

Systematic uncertainties in the
DIPHOX prediction for the shape of the irreducible background 
were obtained by varying the scales of the model and the PDFs, while keeping the overall normalization
fixed in the low mass control region
in which the total background prediction was normalized to the data. The resultant
systematic uncertainties range from a few percent at low masses, up to
$\approx 15$\% for diphoton masses of $\approx 2$~TeV.
Systematic uncertainties in the reducible background shape
were obtained by comparing the results of the extrapolation fit for the various control data
subsamples, in each case maintaining the overall normalization to the
data in the low mass control region.  The resultant uncertainties increase from $\approx 5$\% for low masses
to $\approx 100$\% at a mass of $\approx 2$~TeV.

The systematic uncertainty on the shape of the total background was obtained by adding in quadrature
the uncertainties on the shapes of the irreducible and reducible background components, 
weighted appropriately to account for the purity.  In addition, there is a contribution, which is roughly
constant with a value of $\approx 10$\% for diphoton masses above 800~GeV, introduced 
by varying the purity value within its uncertainty. An additional overall
uncertainty of $\approx 2$\% was included due to the finite statistics of the data sample in
the low mass control region.  

The total background
systematic uncertainty starts at $\approx 2$\% for $m_{\gamma\gamma} = 140$~GeV,
rises to $\approx 15$\% by 700~GeV and then increases slowly up to almost 20\% 
for the highest $m_{\gamma\gamma}$ values, above 2~TeV.

Systematic uncertainties on the signal yields were evaluated separately for the ADD and RS models. 
Since the differences were small, for simplicity the higher value was taken and applied to both models.  
The systematic uncertainties considered for the signal yield
include the 3.7\% uncertainty on the integrated luminosity~\cite{lumi},
and a 1\% uncertainty to account for the limited signal MC statistics. 
A value of 1\% for the uncertainty on the bunch crossing identification (BCID) efficiency accounts for
the ability of the Level 1 trigger hardware to pick the correct BCID when signal pulse saturation occurs in the trigger digitization.  
In addition, a value of 2\% was applied for the uncertainty on the efficiency of the diphoton trigger.  An uncertainty of
2.5\% was applied due to the influence of pileup on the signal efficiency.  Finally, a value of 4.3\% was taken
to account for the uncertainty in the selection and identification of the pair of photons, including uncertainties
due to the photon isolation cut, the description of the detector material, the tight photon identification
requirements, and extrapolation to the high photon $\ET$ values typical of the signal models.  Uncertainties due to
the current knowledge of the EM energy scale and resolution were verified to have a negligible impact.
Adding all effects in quadrature, the total systematic uncertainty on the signal yields was 6.7\%. 

Uncertainties in the theoretical signal cross sections due to PDFs and due to
the NLO approximation were considered. The uncertainties due to PDFs range from
$\approx 10-15$\% for ADD models and from $\approx 5-10$\% for RS models.
The authors of Refs.~\cite{kADD,kRS} have privately updated their calculations
of the NLO signal cross sections for 14~TeV,  and 
provided k-factors to the LHC experiments to scale from LO to NLO
cross section values for the case of 7~TeV $pp$ collisions. 
The NLO k-factor values, evaluated in our
case for $\left| \eta^\gamma \right| < 2.5$, have some modest dependence
on the diphoton mass as well as on $M_S$ for the ADD model,  and on the
 \coupling \ value for the RS model.  However, the variations are within the theoretical uncertainty.
For simplicity, therefore, constant values of 1.70  and 1.75 were assumed for the ADD and RS models, respectively,
and an uncertainty in the k-factor value of $\pm 0.1$ was assigned to account for the variations.

\section{Results and Interpretation}

\begin{figure}[hbpt]\begin{center}
\includegraphics[width=0.49\textwidth]
{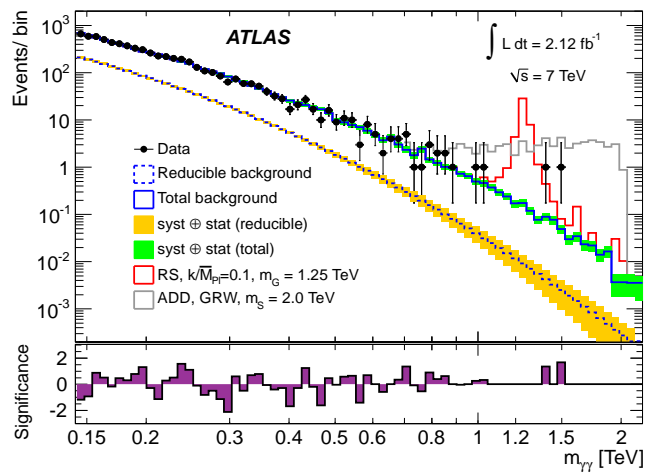}
    \caption{The observed invariant mass distribution of diphoton events, superimposed with the predicted SM background 
    and expected signals for ADD and RS models with certain choices of parameters. The bin width is constant in $\log(m_{\gamma\gamma})$.
    The bin-by-bin significance of the difference between data and background is shown in the lower panel. }
\label{fig:mgg}
\end{center}    
\end{figure}

Figure~\ref{fig:mgg} shows the observed invariant mass distribution of diphoton events, with the predicted 
SM background superimposed as well as ADD and RS signals for certain choices
of the model parameters. The reducible background component 
is shown separately, in addition
to the total background expectation, which sums the reducible and irreducible contributions.  The
shaded bands around each contribution indicate the corresponding uncertainty. The bottom
plot of Figure~\ref{fig:mgg} shows the statistical significance, measured in standard deviations
and based on Poisson distributions, of the difference between
the data and the expected background in each bin. The significance was calculated and displayed
as detailed in Ref.~\cite{significance}, and plotted as positive (negative)
where there was an excess (deficit) in the data in a given bin.  
Table~\ref{tab:events} lists, in bins of diphoton mass, the expected numbers of events for
the irreducible and reducible background components, as well as for the total
background, and also the numbers of observed data events. Both Figure~\ref{fig:mgg} and Table~\ref{tab:events}
demonstrate that there is agreement between the observed
mass distribution and the expectation from the SM backgrounds over the entire diphoton mass range; no evidence
is seen for either resonant or non-resonant deviations which would indicate the presence of a signal due to 
new physics.  An analysis using the BUMPHUNTER~\cite{bumphunter} tool found that the probability, given the
background-only hypothesis, of observing discrepancies at least as large as observed in the data
 was 0.28, indicating quantitatively the good agreement between the data and the
expected SM background.

\begin{table*}[hbpt]\begin{center}
\begin{tabular}{|c|ccc|c|} \hline
Mass Range & \multicolumn{3}{|c|}{Background Expectation} & Observed \\
 (GeV) & Irreducible & Reducible & Total & Events \\ \hline \hline
$[140 , 400]$ &$4738 \pm 180 $ &$1935   \pm 97   $ &$6674   $ & $6674$  \\
$[400 , 500]$ &$90.0 \pm 8.5 $ &$19.9   \pm 1.8  $ &$109.9   \pm 9.2$ & $102$  \\
$[500 , 600]$ &$31.1 \pm 4.0 $ &$5.8    \pm 0.8 $ &$37.0    \pm 4.2$ & $36$  \\
$[600 , 700]$ &$13.7 \pm 2.3 $ &$2.0    \pm 0.4$ &$15.7    \pm 2.4$ & $16$  \\
$[700 , 800]$ &$6.2  \pm 1.2 $ &$0.8    \pm 0.2$ &$6.9     \pm 1.3$ & $9$  \\
$[800 , 900]$ &$3.1  \pm 0.4 $ &$0.3    \pm 0.1$ &$3.4     \pm 0.5$ & $5$  \\
$[900 , 1000]$ &$1.6  \pm 0.2 $ &$0.14   \pm 0.05$ &$1.8     \pm 0.3$ & $1$  \\
$[1000 , 1100]$ &$1.0  \pm 0.2 $ &$0.07   \pm 0.03$ &$1.0     \pm 0.2$ & $1$  \\
$[1100 , 1200]$ &$0.50 \pm 0.09$ &$0.03   \pm 0.02$ &$0.54    \pm 0.11$ & $0$  \\
$[1200 , 1300]$ &$0.29 \pm 0.07$ &$0.02   \pm 0.01$ &$0.31    \pm 0.07$ & $0$  \\
$[1300 , 1400]$ &$0.14 \pm 0.04$ &$0.010  \pm 0.005$ &$0.15    \pm 0.04$ & $1$  \\
$[1400 , 1500]$ &$0.13 \pm 0.04$ &$0.005  \pm 0.003$ &$0.14    \pm 0.04$ & $1$  \\
$ >1500      $ &$0.18 \pm 0.09$ &$0.009  \pm 0.006$ &$0.19    \pm 0.09$ & $0$  \\ \hline
\end{tabular}
\caption{The expected numbers of events for
the irreducible and reducible background components and for the total
background, as well as the numbers of observed data events, in different diphoton mass bins.
The first row, with masses from 140 to 400~GeV, corresponds to the control region in
which the total background was normalized to the corresponding number of observed events.
The errors include both statistical and systematic uncertainties. 
The errors on the irreducible and reducible background
components do not include the contribution, which is anti-correlated between the two background components,  
from the uncertainty on the purity. However, this contribution is included in the errors listed for the total background. }
\label{tab:events}
\end{center}
\end{table*}

Given the absence of evidence for a signal, 95\% CL upper limits were determined on the ADD and RS
signal cross sections,  using a Bayesian approach~\cite{bayes} with a flat
prior on the signal cross section.  The systematic uncertainties 
were incorporated as Gaussian-distributed nuisance parameters and integrated over.
 
To set limits on the ADD model, the number of observed events with diphoton invariant mass in
a high mass signal region was compared with the expected total SM background. 
To optimise the expected limit, the ADD signal search region was chosen 
as $m_{\gamma\gamma} > 1.1$~TeV.  There are 2 observed events in
this signal region, with a background expectation of $1.33 \pm 0.26$ events, where the uncertainty
includes both statistical and systematic errors. The 
observed (expected) $95\%$ CL upper limit is $2.49 \ (1.94)$~fb for the 
product of the cross section due to new physics multiplied by the acceptance and efficiency.
 The cross section result can be translated into limits on $\eta_G$ and, subsequently, on the parameter $M_S$
of the ADD model. As summarized in Table~\ref{tab:ADDlimit}, assuming a k-factor of 1.70,
the 95\% CL lower limits on $M_S$ range between 2.27 and 3.53~TeV,  
depending on the number of extra dimensions assumed and the ADD model implementation.
LO results are also included in Table~\ref{tab:ADDlimit}, for reference.

\begin{table*}[hbpt] \begin{center} 
\begin{tabular}{|c|c|cc|ccccc|} 
\hline 
 k-factor & GRW & \multicolumn{2}{c|}{Hewett} & \multicolumn{5}{c|}{HLZ} \\ 
  Value    &     & Pos & Neg &$n=3$&$n=4$&$n=5$&$n=6$&$n = 7$\\ \hline \hline
 1  & 2.73 & 2.44 &  2.16 & 3.25 & 2.73 & 2.47 & 2.30 & 2.17 \\ \hline
1.70 &  2.97 & 2.66 & 2.27 & 3.53 & 2.97 & 2.69 & 2.50 & 2.36  \\ \hline 
\end{tabular} 
\caption{95\% CL limits on the value of $M_S$ (in TeV) for  various implementations of the ADD model,
using both LO (k-factor = 1) and NLO (k-factor = 1.70) theory cross section calculations.}
\label{tab:ADDlimit}
\end{center} 
\end{table*} 

To determine the limits on the RS model,  the observed invariant mass distribution was compared to templates
of the expected backgrounds and varying amounts of signal for various graviton masses and \coupling \ values.
A likelihood function was defined as the product of the Poisson probabilities over all mass bins in the
search region, defined as $m_{\gamma\gamma} > 500$~GeV, 
where the Poisson probability in each bin was evaluated for the observed number of data
events given the expectation from the template. The total signal acceptance as a function of mass
was propagated into the expectation. The theory uncertainties were
not included in the limit calculation, but are indicated by showing the theory prediction as a band
with a width equal to the combined theory uncertainty when plotting the results.  
The resultant limits are summarized in Table~\ref{tab:RSlimit}.
Using a constant k-factor value of 1.75, the 95\% CL lower limits from
the diphoton channel are $m_G > 0.79 \ (1.85)$~TeV for \coupling \ = 0.01 (0.1).

\begin{figure*}[hbpt]
\begin{center}
\subfigure{\includegraphics[width=0.49\textwidth]{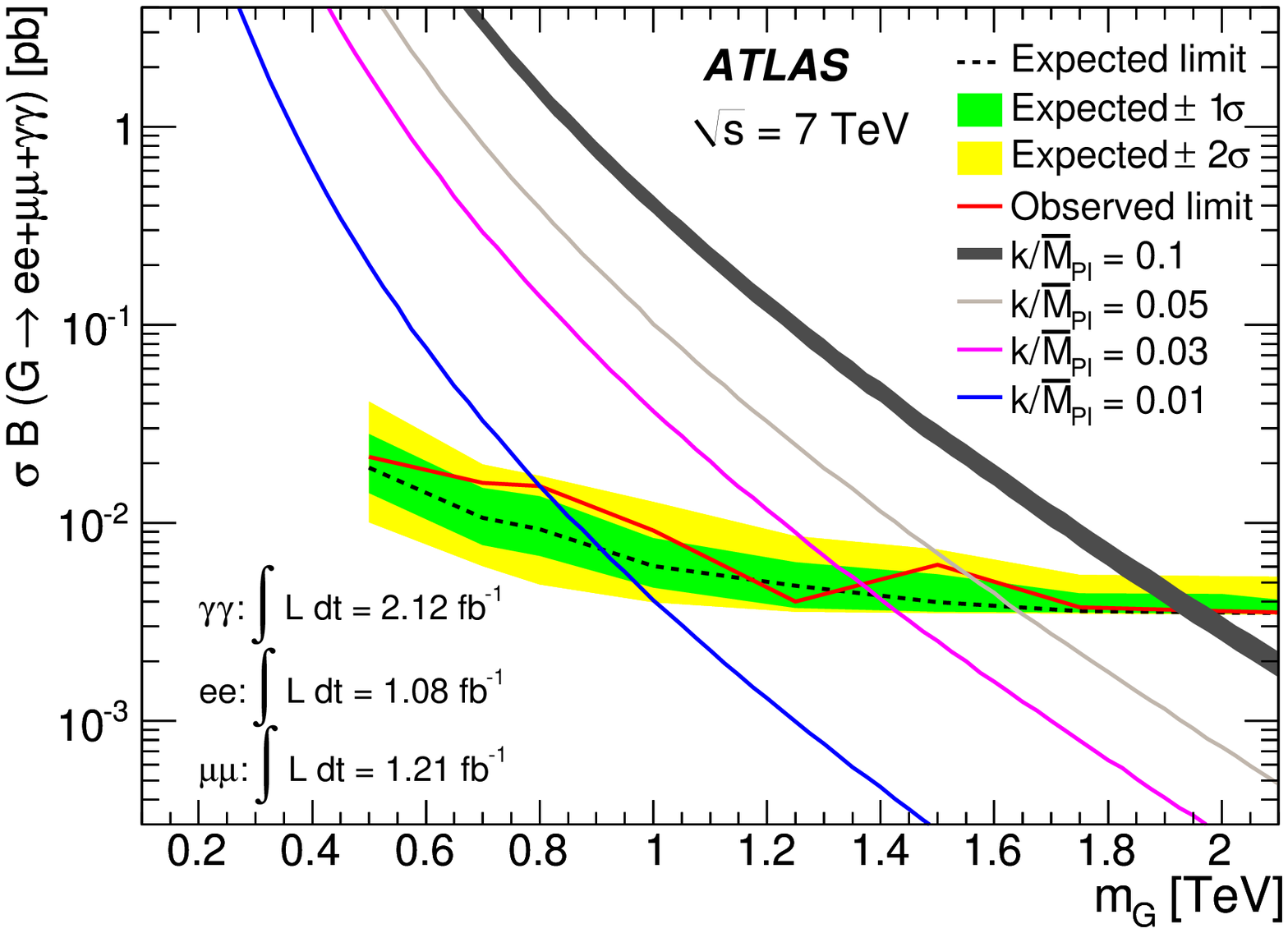}}
 \subfigure{\includegraphics[width=0.49\textwidth]{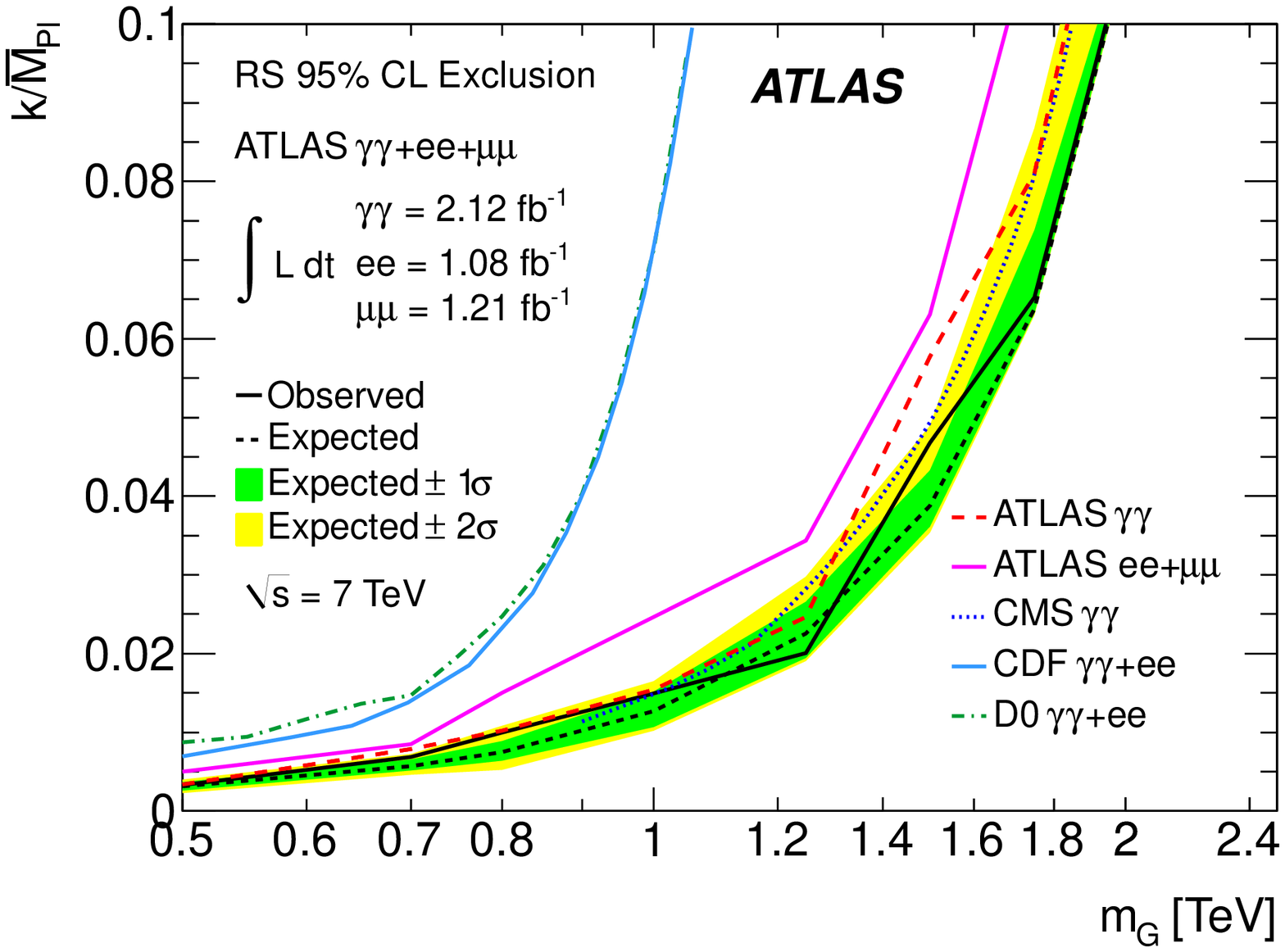}}
\caption{(Left) Expected and observed 95\% CL limits from the combination of $G \rightarrow \gamma\gamma/ee/\mu\mu$
channels on $\sigma B$, the product of  the RS graviton production cross section and
the branching ratio for graviton decay via $G \rightarrow \gamma\gamma/ee/\mu\mu$, as a function of the graviton mass.
The theory curves are drawn assuming a k-factor of 1.75. The thickness of the theory curve for $\coupling = 0.1$ 
illustrates the theoretical uncertainties. (Right)  The RS results interpreted in the plane of \coupling \ versus graviton mass,
and including recent results from other experiments~\cite{TevatronRS,CMSggnew}.
The region above the curve is excluded at 95\% CL. In both figures, linear interpolations are performed
between the discrete set of mass points for which the dilepton limits were calculated in Ref.~\cite{zprime}.}
\label{fig:limits}
\end{center}
\end{figure*}

The RS model results can be combined with the previously published ATLAS results~\cite{zprime} from the
dilepton final state, where, assuming LO cross
sections and $\coupling = 0.1$, RS gravitons with masses below 1.51 (1.45)~TeV were excluded
at 95\% CL using data samples of 1.08 (1.21)~fb$^{-1}$ to search for 
$G \rightarrow ee$ ($G \rightarrow \mu\mu$). To ensure their statistical independence,
the selection cuts of the diphoton analysis included a veto of any events
which were also selected by the 1.08~fb$^{-1}$ $G \rightarrow ee$ analysis.  
In performing the combination, correlations were considered between the systematic uncertainties
in the $\gamma\gamma$ and $ee$ channels. In the $ee$ analysis~\cite{zprime}, the background
prediction was normalized such that the expected and observed numbers of events in the region 
of the $Z$ peak agreed, eliminating the dependence of the $ee$ result on the
measured integrated luminosity. Therefore, the $\gamma\gamma$ and $ee$ signal predictions were treated as uncorrelated,
since there should be no correlation in the luminosity and efficiency uncertainties. The systematic uncertainty on the 
QCD dijet background was treated as being correlated; however, this background was quite small
so the effect was minor. The PDF and scale uncertainties were treated as correlated across all three channels, and
affect the irreducible background in the $\gamma\gamma$ channel as well as the Drell-Yan background
in the $ee/\mu\mu$ channels. The left plot of Figure~\ref{fig:limits} shows the combined 95\% CL upper limit on the  
product of the graviton production cross section times the branching ratio for 
$G \rightarrow \gamma\gamma/ee/\mu\mu$, obtained using the same k-factor value
of 1.75 for all three channels.  As summarized in Table~\ref{tab:RSlimit}, the combined 95\% CL lower limit 
is $m_G > 0.80 \ (1.95)$~TeV for \coupling \ = 0.01 (0.1).
As shown in the right plot of Figure~\ref{fig:limits}, the results can be translated 
into a 95\% CL exclusion in the plane of \coupling \ versus graviton mass.

\begin{table}[hbpt]\begin{center}
    \begin{tabular}{|c|c| c c c c|} \hline
      \multirow{3}{45pt}{\centering{k-Factor Value}}  & \multirow{3}{60pt}{\centering{Channel(s) Used}} & \multicolumn{4}{c|}{95\% CL Limit [TeV]} \\ \cline{3-6}
      &                                  & \multicolumn{4}{c|}{ $k/\overline{M}_{Pl}$ Value}  \\
      &                                             & 0.01     &0.03      &0.05        &0.1            \\ \hline\hline
      \multirow{2}{*}{1}     &$G\to\gamma\gamma$    &0.74      &1.26      &1.41        &1.79  \\
      & $G\to \gamma\gamma$/$ee$/$\mu\mu$           &0.76      &1.32      &1.47        &1.90  \\ \hline \hline
      \multirow{2}{*}{1.75} & $G\to\gamma\gamma$    &0.79      &1.30      &1.45        &1.85  \\
      &$G\to \gamma\gamma$/$ee$/$\mu\mu$            &0.80      &1.37      &1.55        &1.95  \\\hline
    \end{tabular}
\caption{95\% CL lower limits on the mass (GeV) of the lightest RS graviton, for  various
values of \coupling. The results are shown for the diphoton channel alone and for
the combination of the diphoton channel with the dilepton results of Ref.~\cite{zprime}, using both LO (k-factor = 1) and
NLO (k-factor = 1.75) theory cross section calculations.}
\label{tab:RSlimit}
\end{center}
\end{table}  

\section{Summary}

Using a dataset corresponding to 2.12~fb$^{-1}$, an analysis of the 
diphoton final state was used to set 95\% CL lower limits of between 2.27 and 3.53~TeV on the parameter 
$M_S$  of the ADD large extra dimension scenario, depending on the number of extra dimensions and the theoretical formalism used. 
The diphoton results also exclude at 95\% CL RS graviton masses below 0.79 (1.85)~TeV for the dimensionless RS coupling \coupling = 0.01 (0.1).
Combining with the previous ATLAS dilepton analyses further tightens these limits to exclude at 95\% CL RS graviton masses 
below 0.80 (1.95)~TeV for  \coupling = 0.01 (0.1). 

\vspace{0.2cm}



\section{Acknowledgements}

We thank CERN for the very successful operation of the LHC, as well as the
support staff from our institutions without whom ATLAS could not be
operated efficiently.

We acknowledge the support of ANPCyT, Argentina; YerPhI, Armenia; ARC,
Australia; BMWF, Austria; ANAS, Azerbaijan; SSTC, Belarus; CNPq and FAPESP,
Brazil; NSERC, NRC and CFI, Canada; CERN; CONICYT, Chile; CAS, MOST and
NSFC, China; COLCIENCIAS, Colombia; MSMT CR, MPO CR and VSC CR, Czech
Republic; DNRF, DNSRC and Lundbeck Foundation, Denmark; ARTEMIS, European
Union; IN2P3-CNRS, CEA-DSM/IRFU, France; GNAS, Georgia; BMBF, DFG, HGF, MPG
and AvH Foundation, Germany; GSRT, Greece; ISF, MINERVA, GIF, DIP and
Benoziyo Center, Israel; INFN, Italy; MEXT and JSPS, Japan; CNRST, Morocco;
FOM and NWO, Netherlands; RCN, Norway; MNiSW, Poland; GRICES and FCT,
Portugal; MERYS (MECTS), Romania; MES of Russia and ROSATOM, Russian
Federation; JINR; MSTD, Serbia; MSSR, Slovakia; ARRS and MVZT, Slovenia;
DST/NRF, South Africa; MICINN, Spain; SRC and Wallenberg Foundation,
Sweden; SER, SNSF and Cantons of Bern and Geneva, Switzerland; NSC, Taiwan;
TAEK, Turkey; STFC, the Royal Society and Leverhulme Trust, United Kingdom;
DOE and NSF, United States of America.

The crucial computing support from all WLCG partners is acknowledged
gratefully, in particular from CERN and the ATLAS Tier-1 facilities at
TRIUMF (Canada), NDGF (Denmark, Norway, Sweden), CC-IN2P3 (France),
KIT/GridKA (Germany), INFN-CNAF (Italy), NL-T1 (Netherlands), PIC (Spain),
ASGC (Taiwan), RAL (UK) and BNL (USA) and in the Tier-2 facilities
worldwide.


\clearpage
\onecolumn
\begin{flushleft}
{\Large The ATLAS Collaboration}

\bigskip

G.~Aad$^{\rm 48}$,
B.~Abbott$^{\rm 110}$,
J.~Abdallah$^{\rm 11}$,
A.A.~Abdelalim$^{\rm 49}$,
A.~Abdesselam$^{\rm 117}$,
O.~Abdinov$^{\rm 10}$,
B.~Abi$^{\rm 111}$,
M.~Abolins$^{\rm 87}$,
O.S.~AbouZeid$^{\rm 157}$,
H.~Abramowicz$^{\rm 152}$
% ...
\bigskip

% ...
$^{10}$ Institute of Physics, Azerbaijan Academy of Sciences, Baku, Azerbaijan\\
$^{11}$ Institut de F\'isica d'Altes Energies and Departament de F\'isica de la Universitat Aut\`onoma  de Barcelona and ICREA, Barcelona, Spain\\
% ...
$^{48}$ Fakult\"{a}t f\"{u}r Mathematik und Physik, Albert-Ludwigs-Universit\"{a}t, Freiburg i.Br., Germany\\
$^{49}$ Section de Physique, Universit\'e de Gen\`eve, Geneva, Switzerland\\
% ...
$^{87}$ Department of Physics and Astronomy, Michigan State University, East Lansing MI, United States of America\\
% ...
$^{110}$ Homer L. Dodge Department of Physics and Astronomy, University of Oklahoma, Norman OK, United States of America\\
$^{111}$ Department of Physics, Oklahoma State University, Stillwater OK, United States of America\\
% ...
$^{117}$ Department of Physics, Oxford University, Oxford, United Kingdom\\
% ...
$^{152}$ Raymond and Beverly Sackler School of Physics and Astronomy, Tel Aviv University, Tel Aviv, Israel\\
% ...
$^{157}$ Department of Physics, University of Toronto, Toronto ON, Canada\\
% ...
\end{flushleft}



\section*{References}
\begin{thebibliography}{99}

\bibitem{ADD}
N.~Arkani-Hamed, S. Dimopoulos, and G.R. Dvali, Phys. Lett. {\bf B429}, 263 (1998).

\bibitem{RSmodel} L. Randall and R. Sundrum, Phys. Rev. Lett. {\bf 83}, 3370 (1999).

\bibitem{GRW}
G. Giudice, R. Rattazzi, and J. Wells, Nucl. Phys. {\bf B544}, 3 (1999).

\bibitem{HLZ}
T. Han, J. Lykken, and R.-J. Zhang, Phys. Rev. {\bf D59}, 105006 (1999).

\bibitem{Hewett}
J. Hewett, Phys. Rev. Lett. {\bf 82}, 4765 (1999).

\bibitem{HERAADD}
H1 Collaboration, Phys. Lett.  {\bf B568}, 35 (2003);
ZEUS Collaboration, Phys. Lett. {\bf B591}, 23 (2004).

\bibitem{LEPADD}
LEP Working Group LEP2FF/02-02 (2002);
LEP Working Group LEP2FF/03-01 (2003);
ALEPH Collaboration, Eur. Phys. J. {\bf C49}, 411 (2007).

\bibitem{TevatronADD}
D0 Collaboration, Phys. Rev. Lett. {\bf 102}, 051601 (2009);
D0 Collaboration, Phys. Rev. Lett. {\bf  103}, 191803 (2009).

\bibitem{CMSADD}
CMS Collaboration, JHEP {\bf 1105}, 085 (2011).

\bibitem{CMSggnew}
CMS Collaboration, arXiv:1112.0688 (2011) (submitted to Phys. Rev. Lett.).

\bibitem{coupling}
H. Davoudiasl, J. L. Hewett and T. G. Rizzo, Phys. Rev. Lett. {\bf 84}, 2080 (2000).

\bibitem{zprime}
ATLAS Collaboration, arXiv:1108.1582 (2011) (accepted by Phys. Rev. Lett.).

\bibitem{TevatronRS}
D0 Collaboration, Phys. Rev. Lett. {\bf 104}, 241802 (2010);
CDF Collaboration, Phys. Rev. Lett. {\bf 107}, 051801 (2011).

\bibitem{CMSRSeemm}
CMS Collaboration, JHEP {\bf 1105}, 093 (2011).

\bibitem{DetectorPaper}
ATLAS Collaboration, JINST {\bf 3}, S08003 (2008).

\bibitem{ATLASphotons}
ATLAS Collaboration, Phys. Rev. {\bf D83}, 052005 (2011).

\bibitem{ambientenergy}
M. Cacciari, G. P. Salam, and S. Sapeta, JHEP {\bf 04}, 065 (2010).

\bibitem{ATLASsim}
ATLAS Collaboration, Eur. Phys. J. {\bf C70} (2010) 823.

\bibitem{geant4}
S. Agostinelli {\sl et al.} (GEANT4 Collaboration), Nucl. Instrum. Methods {\bf A506}, 250 (2003).

\bibitem{ATLAStune}
ATLAS Collaboration, ATL-PHYS-PUB-2010-014 (2010), \url{https://cdsweb.cern.ch/record/1303025}.

\bibitem{pythia}
T. Sj\"{o}strand {\sl et al.}, Comput. Phys. Commun. {\bf 135}, 238 (2001).

\bibitem{MRST2007lomod}
A. Sherstnev and R. S. Thorne, Eur. Phys. J. {\bf C55}, 553 (2008).

\bibitem{diphox}
T. Binoth, {\sl et al.}, Eur. Phys. J. {\bf C16}, 311 (2000). 

\bibitem{pdf4lhc}
PDF4LHC Working Group, arXiv:1101.0536 (2011).
 
\bibitem{mstw2008}
A. D. Martin, {\sl et al.}, Eur. Phys. J. {\bf C63}, 189 (2009).

\bibitem{cteq66}
P. M. Nadolsky, {\sl et al.}, Phys. Rev. {\bf D78}, 013004 (2008).

\bibitem{sherpa} 
T. Gleisberg et al., JHEP {\bf 02}, 007 (2009).

\bibitem{egamma}
ATLAS Collaboration, arXiv:1110.3174 (2011), submitted to EPJC.

\bibitem{ATLASdijet}
ATLAS Collaboration, arXiv:1108.6311 (2011); submitted to Phys. Lett. B.

\bibitem{Hgg}
ATLAS Collaboration, arXiv:1108.5895 (2011), submitted to Phys. Lett. B.

\bibitem{SMdiphoton}
ATLAS Collaboration, arXiv:1107.0581 (2011); submitted to Phys. Rev. D.

\bibitem{lumi}
ATLAS Collaboration, Eur. Phys. J. {\bf C71}, 1630 (2011);
ATLAS Collaboration, ATLAS-CONF-2011-116 (2011), \\
\url{https://cdsweb.cern.ch/record/1376384}.

\bibitem{kADD}
M. Kumar, {\sl et al.},  Phys. Lett. {\bf B672}, 45 (2009).

\bibitem{kRS}
M. Kumar, {\sl et al.}, arXiv:0902.4894  (2009).

\bibitem{significance}
G. Choudalakis and D. Casadei, arXiv:1111.2062 (2011).

\bibitem{bumphunter}
CDF Collaboration, Phys. Rev. {\bf D79}, 011101 (2009);
G. Choudalakis, arXiv:1101.0390 (2011).

\bibitem{bayes}
A. Caldwell, D. Kollar, and K. Kr\"oninger, Comput. Phys. Commun. {\bf 180}, 2197 (2009).

\end{thebibliography}
\end{document}